\documentclass[
aps, pra,
amsmath,amssymb,
11pt,
final,
tightenlines,
twoside,
twocolumn,
nofloats,
nofootinbib,
superscriptaddress,
showkeys,
showkeywords,
]
{revtex4-2}

\usepackage[T2A]{fontenc}
\usepackage[utf8x]{inputenc}
\usepackage[russian,english]{babel}
\usepackage{graphicx}
\usepackage{dcolumn}
\usepackage{bm}

\input{maik.rty}

\makeatletter
\def\@keys@name{Keywords: }
\makeatother

\setcitestyle{authoryear,round,aysep={,},citesep={;}}
\PassOptionsToPackage{authoryear}{natbib}
\def\squareforqed{\hbox{\rlap{$\sqcap$}$\sqcup$}}

\def\sq{\ifmmode\squareforqed\else{\unskip\nobreak\hfil
\penalty50\hskip1em\null\nobreak\hfil\squareforqed
\parfillskip=0pt\finalhyphendemerits=0\endgraf}\fi}

\def\utw{\smash{\rlap{\lower5pt\hbox{$\sim$}}}}

\def\udtw{\smash{\rlap{\lower6pt\hbox{$\approx$}}}}

\def\fs{\hbox{$\,.\!\!^{\rm s}$}}

\def\farcs{\hbox{$\,.\!\!^{\prime\prime}$}}

\def\diameter{{\ifmmode\mathchoice
{\ooalign{\hfil\hbox{$\displaystyle/$}\hfil\crcr
{\hbox{$\displaystyle\mathchar"20D$}}}}
{\ooalign{\hfil\hbox{$\textstyle/$}\hfil\crcr
{\hbox{$\textstyle\mathchar"20D$}}}}
{\ooalign{\hfil\hbox{$\scriptstyle/$}\hfil\crcr
{\hbox{$\scriptstyle\mathchar"20D$}}}}
{\ooalign{\hfil\hbox{$\scriptscriptstyle/$}\hfil\crcr
{\hbox{$\scriptscriptstyle\mathchar"20D$}}}}
\else{\ooalign{\hfil/\hfil\crcr\mathhexbox20D}}%
\fi}}

\def\*{$^{*}$}

\DeclareRobustCommand{\ion}[2]{%
    \relax\ifmmode
    \ifx\testbx\f
    {\mathrm{#1\,\textsc{#2}}}\else
    {\mathrm{#1\,\mathsc{#2}}}\fi
    \else\textup{#1\,{\mdseries\textsc{#2}}}%
    \fi}
\makeatletter
\def\p@subsection{}
\def\p@subsubsection{}
\makeatother

\begin{document}

\selectlanguage{english}
\title{TWO FAST RADIO BURST CANDIDATES DETECTED IN THE RATAN-600 BLIND SURVEYS}

\author{\firstname{S.~A.}~\surname{Trushkin}}\email{satr@sao.ru} 
\affiliation{Special Astrophysical Observatory,  Russian Academy of Sciences,
Nizhnii Arkhyz, 369167 Russia}

\author{\firstname{D.~O.}~\surname{Kudryavtsev}}
\affiliation{Special Astrophysical Observatory,  Russian Academy of Sciences,
Nizhnii Arkhyz, 369167 Russia}

\author{\firstname{P.~G.}~\surname{Tsybulev}}
\affiliation{Special Astrophysical Observatory,  Russian Academy of Sciences,
Nizhnii Arkhyz, 369167 Russia}

\author{\firstname{N.~N.}~\surname{Bursov}}
\affiliation{Special Astrophysical Observatory,  Russian Academy of Sciences,
Nizhnii Arkhyz, 369167 Russia}

\begin{abstract}
We report the detection of two single dispersed pulses with widths of about 6.7\,ms and 15.6\,ms and flux densities of about 0.8 and 1.7\,Jy. These were found in a blind survey of two circular regions (Dec$\;=+22^{\circ}$ and $+5^{\circ}$, RA: 0--24~h) at high radio frequencies \mbox{(4400--5000\,MHz)} using four 4-channel filter radiometers (central frequencies of 4475, 4625, 4775, and 4925\,MHz) spanning  a total frequency bandwidth of 600\,MHz. The survey was carried out with the Western Sector antenna of the RATAN-600 radio telescope, which had an effective area of 1200\,m$^2$ in~2017--2025. The total area of the survey was about 1500~square degrees, and each pointing of the antenna lasted one year. We consider these to be two rare,  high-frequency FRB events (MJD~58614.38257 and MJD~58659.62482), detected with estimated dispersion measures of $200\pm20$ and $266\pm10$\,pc\,cm$^{-3}$  at a final signal-to-noise ratio (SNR) level of 8--9 after additional optimal filtration. After subtraction of the estimated dispersions within the Milky Way, we estimated redshifts of approximately 0.09 and 0.13, respectively, using the traditional equation:  ${\rm DM}\,[{\rm pc~cm}^{-3}] = 1000\,z$ for the intergalactic medium.  The estimated isotropic energies of the FRBs are equal to $10^{40}$\,erg and $2\times 10^{40}$\,erg, respectively. To effectively detect single pulses, we applied machine learning methods using a neural network trained on  synthetic FRB pulse profiles.
\end{abstract}

\maketitle

\section{INTRODUCTION}

Fast radio bursts (FRBs) have become the most intriguing cosmic radio sources over the past decade \citep{2019A&ARv..27....4P}. 
While the discovery of the single ``Lorimer burst'' (now FRB\,010724) in 2007 \citep{2007Sci...318..777L} 
was met with some skepticism, subsequent discoveries in 2013 and later firmly established them as a reality. During this period, pulses with durations from 1 to 20~ms and peak flux densities of several Jy were discovered with the Parkes \mbox{64-m} telescope \citep{2013Sci...341...53T} 
and other instruments using the well-known ``pulsar machines.\!'' It then became clear that we were dealing with a new class of pulsed radio emission sources. Furthermore, their extragalactic origin was confirmed since these FRBs showed large dispersion measures (DM) even at high Galactic latitudes.

In 2017, when we started our research, the number of known FRBs did not exceed 50, and most of them had been detected in the sky surveys with the 64-m telescope in Parkes (Australia). It was clear, however, that all major telescopes were increasingly joining the search for new events. Especially impressive were the studies of the repeating FRB\,121102. Originally discovered with the 300-m Arecibo telescope (USA), it was subsequently monitored by the JVLA, the \mbox{100-m} telescope in Effelsberg (Germany) \citep{2017ApJ...850...76L}, 
and recently by FAST (China) \citep{2026ATel17642....1Z}. 
Radio waves from FRB\,121102 had been recorded across a broad frequency range, from 400\,MHz to 5\,GHz. Later, other FRBs were detected even at 144\,MHz with LOFAR \citep{2021ApJ...911L...3P} 
and at 111\,MHz with the LPA (Puschino, LPI~RAS) \citep{2025PASA...42...59T}. 

Therefore, when choosing our search frequency, we settled on 4700\,MHz. We reasonably assumed that a favorable interference situation, the high sensitivity of HEMT amplifiers in this band, the low noise temperature of the antenna, and the need for only a small number of filter channels---necessary to search for pulses subject to dispersion---offered strong prospects for a successful survey  \citep{2018sn87.conf..211T}. 

To avoid interfering with the intensive observing
programs on the RATAN-600 sectors equipped with an antenna control system which allows for up to 100~observations per day, we chose the Western Sector. This antenna can occasionally be adjusted to different elevations in the direction of the East (Azimuth~$90^{\circ}$). Four four-channel radiometers, each with a 150\,MHz bandwidth and providing a total bandwidth of 600\,MHz, were developed and purchased by SAO~RAS; they were installed at the focus of secondary mirror No.~5.

Since the antenna is stationary, during a survey a given point in the sky passes across each of the four beams in 5--8~s (at half the beam width), meaning the total ``standstill'' time for such a point is $6~\rm s\times 4~s\times 365\,days = 8760\,s = 146$~minutes. This is not the highest figure for blind surveys; many repeating FRBs have been observed for several hours at a time. For example, a 40-hour tracking of the ``Lorimer burst''\ has reasonably showed that it is not a repeating FRB.

During the first year (from mid-2017 to mid-2018), the antenna was set to the declination of \mbox{FRB 121102}. However, we were unable to detect any bursts, although later epochs of heightened activity demonstrated a burst rate exceeding 400~per hour 
\citep{2026ATel17642....1Z}. Since~2017, we have been conducting surveys at various declinations, selecting declinations of radio sources that could be monitored along the way. In addition to blind FRB searches, this survey approach could yield new interesting results for discrete sources. Indeed, in the survey at the elevation of the Crab Nebula pulsar, we recorded several giant pulses in one of the beams, which were, as expected, shifted in time by the known dispersion measure of~57\,pc\,cm$^{-3}$. The survey at this elevation from June~2018 to June~2019 provided intensive monitoring of 205~extragalactic sources \citep{2024AstBu..79...36K}.
Furthermore, a survey at the elevation of the microquasar SS\,433 showed that in 95\%\ of the cases, the flux densities at 4.7\,GHz coincided within the uncertainties with the measurements obtained using the Northern Sector of \mbox{RATAN-600}, despite a time difference of about 5.5~hours between the observations. This indicates that the characteristic variability timescale of the radio flux from the SS\,433 jets is longer than five hours. In this paper we discuss the detection of two FRBs in just two surveys at declinations $\rm{Dec}_{\rm{J2000}} = +22^{\circ}$ (the Crab pulsar elevation) and $\rm{Dec}_{\rm{J2000}} = +5^{\circ}$ (SS\,433).  The search for bursts using the same technique was carried out across nine annual surveys, covering a total celestial area of about 20\,000~square degrees. On average, we sampled each ``pixel'' ($1^\prime\times 36^\prime$) of the sky section 1500~times annually, accumulating a total of 30\,TB of data per year at a sampling interval of approximately 0.245\,ms.

Given that the sensitivity of a single receiver (beam) at 4.7\,GHz over the broad bandwidth is 5~mJy, we can roughly estimate the sensitivity within a single narrow channel for a 10-ms-wide burst as:
\begin{equation}
\label{eq:time_delay}
\begin{split}
\rm 5~mJy\times\left(\frac{6~s}{0.01~s}\right)^{0.5}\left(\frac{600~MHz}{150~MHz}\right)^{0.5}\!\!=245~mJy. \nonumber
\end{split}
\end{equation}
This corresponds to a sensitivity level of approximately 3\,Jy at a reference frequency of 1\,GHz.

The RATAN-600 data obtained in the blind surveys with the Western Sector antenna consist of four-channel time series. Each subchannel  records the signal in units of antenna temperature across a 150\,MHz band centered at frequencies of 4475, 4625, 4775, and 4925\,MHz within the total bandwidth of 4400--5000\,MHz.  As radio waves propagate through the ionized interstellar and intergalactic media, their group velocity falls below the speed of light in the vacuum. The pulse travels more slowly at lower frequencies than at higher frequencies. Consequently, a short pulse emitted by a distant radio source experiences a frequency-dependent time delay due to cosmic dispersion in both the Galactic and extragalactic media (see \citealt{2012hpa..book.....L}): 
\begin{equation}
\label{eq:time_delay}
\begin{split}
\Delta t [\rm ms] & =\cfrac{e^2}{2\pi m_e c}\, {\rm DM}\left(\frac{1}{\nu_1^2}- \cfrac{1}{\nu_2^2}\right) \\
	      &= 4.1488\,{\rm DM}\left(\cfrac{1}{\nu_1^2}-\cfrac{1}{\nu_2^2}\right),
\end{split}
\end{equation}
where $\Delta t$ is the time delay (in ms) between  two frequencies $\nu_1<\nu_2$ in GHz, $\rm DM$ is the dispersion measure in pc\,cm$^{-3}$, $e$ and $m_e$ are the electron charge and mass, and $c$ is the speed of light. These delays in the arrival times of pulses from pulsars can be used to identify real cosmic signals under the conditions of radio frequency interference (RFI) originating in the near-Earth environment. For our frequency band, the delays are of the order of a few milliseconds between the subchannels for $\rm DM=100\,pc\,cm^{-3}$. The broadening of millisecond pulses due to scattering in the interstellar medium is  negligibly small at 4.7\,GHz. Nevertheless, the in-band (150\,MHz) broadening can be notable (3\,ms) for $\rm DM> 100$\,pc\,cm$^{-3}$, according to the formula $W_{\rm DM}~\rm[ms]=8.3\times{\rm DM}\,\Delta\nu \times \nu^{-3}$, where $\Delta\nu$ and $\nu$ are given in GHz.

Impulsive interference, which is especially frequent during the day, is the main problem in searching for FRBs. However, it can be effectively eliminated by comparing the times of arrival (TOA) of the pulse in the four channels: if there is no shift between the signals, it is clearly interference. This first labor-intensive stage of sifting out most of the RFI signals was accomplished using convolutional neural networks.

\section{DETECTION TECHNIQUE}

The detection technique is described in detail in our companion paper 
\citet{2026A&C....5401002K}; here, we only briefly outline the principal points of the procedure. To detect rare events such as FRBs in long-term drift scans, we employed an approach based on convolutional neural networks (CNNs). Although several neural networks dedicated to FRB searches have previously been developed (e.g., \citealt{2018AJ....156..256C,2018ApJ...866..149Z, 2020MNRAS.497.1661A}) 
the RATAN-600 data exhibit distinct characteristics. Most notably, our data contain only four channels, as opposed to the hundreds or thousands available on other radio telescopes. Therefore, we developed a custom network architecture based on the Efficient\,Net family of models \citep{pmlr-v97-tan19a,pmlr-v139-tan21a} 
Our neural network is optimized for operating with multichannel time series using 1D convolutions, which are well-suited for processing non-2D input formats and effectively reduce the number of trainable parameters. The primary objective of the neural network is to distinguish a genuine FRB pulse from short-duration RFI signals or from the radiometer's white and and $1/f$ noises.

During inference, the drift scan records---each spanning approximately one sidereal hour with a sampling interval of about 245\,$\mu$\,s---are segmented into 1-s intervals and processed by a neural network for classification 
\citep{2026A&C....5401002K}.
In general, the neural network can be considered a computational graph that takes a radiometer record as input and provides the probability estimate of a signal (an event) being present. The network is trained in such a way that the 50\% probability estimate corresponds to the separation between the classes in the training sample. We should note that this value is not a true mathematical probability, but it could be loosely referred to as the confidence of the network in the presence of a signal. Although the network has shown low confidence in the detection of the candidate events considered in this paper (11--14\%),  they can be further processed using a more detailed analysis including suppression of high-frequency noise. As is shown further in Section~3, we have obtained signal-to noise ratios of 8--9 for the pulses, which is considered a significant detection ($\rm SNR >7$). On the other hand, given the huge number of the instances analyzed in the surveys (750 billion \mbox{1\,s-intervals}), we cannot fully exclude an accidental combination of weak RFI pulses fluctuations that mimic an FRB.

To minimize false-positive detections, a cascade of two independently trained neural networks, optimized probability thresholds, and straightforward algorithmic filtering were applied  
\citep{2026A&C....5401002K}. The final candidates were inspected visually.

\subsection{Event Modeling and Estimation of Pulse~Parameters}
\label{sec:event_modeling}

While bright, high-amplitude pulses can be readily identified in raw records, typical events exhibit low amplitudes and require specialized processing to be distinguished from the noisy background. Neural networks isolate candidates based on statistical deviations that resemble the features presented in the training dataset. However, the subsequent stage requires validating these presumed true-positive detections using more detailed and sophisticated analysis. In our case, the limited number of frequency channels introduces unique challenges to this procedure.

If the dispersion measure ($\rm DM$), which relates to the observed time delay via Eq.~(\ref{eq:time_delay}), is known, the observed time profiles in distinct radio channels can be dedispersed and co-added to obtain the total time profile of the pulse. From this integrated profile, other key characteristics can be derived, including the pulse width, peak antenna temperature (or flux density), and fluence (the time-integrated flux density). The standard technique for estimating the observed $\rm DM$ involves iterating over a grid of $\rm DM$ values to maximize the signal-to-noise ratio ($\rm SNR$) of the co-added profile. In practice, a pulse is represented as an image in the $\rm DM$--time domain (Fig.~1), where the maximum $\rm SNR$ is typically located by thresholding the intensity and applying, for example, a 2D Gaussian fit.

With a small number of channels, however, the localization of the high-intensity region along the $\rm DM$ axis is inherently poor. Therefore, rather than iterating through $\rm DM$ values to search for the peak $\rm SNR$, we determine the pulse parameters by directly fitting the observed per-channel profiles with a physical best-fit model.

\begin{figure*}[ht!]
\includegraphics[width=0.8\linewidth]{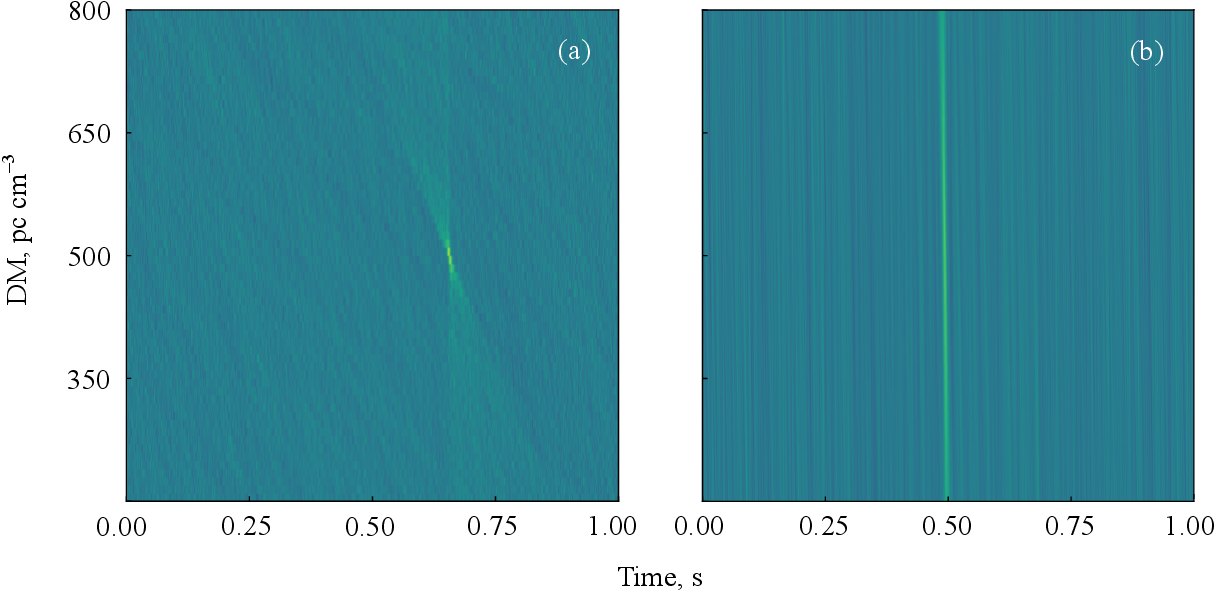}
\caption{Comparison of synthetic $\rm DM$--time diagrams for a system with a large number of channels (a) and the four-channel configuration of RATAN-600 (b). Despite the reduced channel count, pulse parameters can still be effectively constrained using only four channels.}
\label{fig:dmt_compar}
\end{figure*}

The procedure consists of constructing model FRB pulses and optimizing their parameters to minimize the residuals between the model and the observed time profiles. The following parameters are fitted simultaneously: $\rm DM$, pulse arrival time, width, and per-channel amplitudes. A model FRB pulse for a given set of parameters is calculated following the methodology described in 
\citet{2026A&C....5401002K}. The intrinsic time profile is assumed to be Gaussian, with scattering and in-band dispersion smearing additionally taken into account. It is worth noting that simplifying the model pulse profile in this manner does not allow us to include into the fitting the possible presence of an internal structure in the profiles of real fast radio bursts; however, the structure of an observed time profile can be preserved after fitting the general Gaussian-like shape.

The time delays are calculated according to Eq.~(1). The pulse arrival time is defined as the center of the Gaussian profile in the highest-frequency channel. Hereinafter, the pulse width refers to the full width at half maximum (FWHM) of the model Gaussian. The amplitudes of the time profiles across different channels are susceptible to random nulling effects and, therefore, must be optimized as independent parameters. While scattering could potentially be introduced as an additional free parameter, its effect is negligible at the high radio frequencies (4.7\,GHz), and thus it was not optimized in this study.

The optimal parameter values, i.e., the observed parameters of the pulse, are found based on the Bayesian inference using the nested sampling Monte Carlo algorithm {\tt MLFriends} 
\citep{2019PASP..131j8005B} implemented in the {\tt UltraNest}\footnote{\url{https://johannesbuchner.github.io/UltraNest/}} package \citep{2021JOSS....6.3001B}. 
The likelihood was taken as
\begin{equation}
\mathcal{L}(\boldsymbol{\theta}|S(t)) = -\frac{1}{2}\sum_{\rm ch=1}^{4}\sum_{t\in\mathcal{T}}\left(\frac{\hat{S}(t|\boldsymbol{\theta})-S(t)}{\sigma_{\rm noise}}\right)^2,\
\end{equation}
where $\boldsymbol{\theta}$ represents the vector of the desired free parameters, $S(t)$ are the observed time profiles, $\hat{S}(t|\boldsymbol{\theta})$ are the model profiles calculated as described above, $\sigma_{\rm noise}$ is the standard deviation of radiometer noise, and the summation is performed over all frequency channels within the pulse duration $\mathcal{T}$. With this likelihood, the optimization problem is mathematically equivalent to a weighted least-squares fitting. However, the Bayesian inference accommodates complex models and inherently provides robust uncertainty estimates for the optimized parameters.

\subsection{Reduction of High-Frequency Noise}

For low-amplitude FRB pulses, the radiometer noise can significantly degrade the observed time profiles. In observations conducted with a large number of frequency channels, an FRB pulse profile may be entirely invisible in any individual channel and can only be recovered as an integrated profile after dedispersion and co-addition of the subchannels (e.g., see Fig.~3 in \citealt{2018ApJ...866..149Z}). 
In contrast, a system with few channels faces severe constraints because the amplitude and SNR of the integrated profile do not increase substantially after summation: by a factor of two at best for our four channels, assuming the absence of scintillations. Nevertheless, the neural network is highly sensitive to the specific statistical deviations characteristic of FRB pulses, enabling the detection of candidates with low amplitudes comparable to the noise level. To verify these candidate events, we apply a low-pass Butterworth filter, which suppresses high-frequency temporal noise while preserving the lower-frequency pulse profile intact.

Figure~2a illustrates a synthetic time profile injected with white noise at an SNR of 1.5, matching the typical levels observed in our candidate events. The bottom panel displays the result of applying an 8th-order low-pass Butterworth filter. The pulse is effectively recovered, achieving a noise reduction by a factor of approximately ten. In real observations, this filtering is somewhat less effective due to the presence of the $1/f$ noise component in the radiometer output alongside the white noise; nonetheless, the procedure remains highly efficient, as demonstrated below.
\begin{figure*}[ht!]
\includegraphics[width=0.76\linewidth]{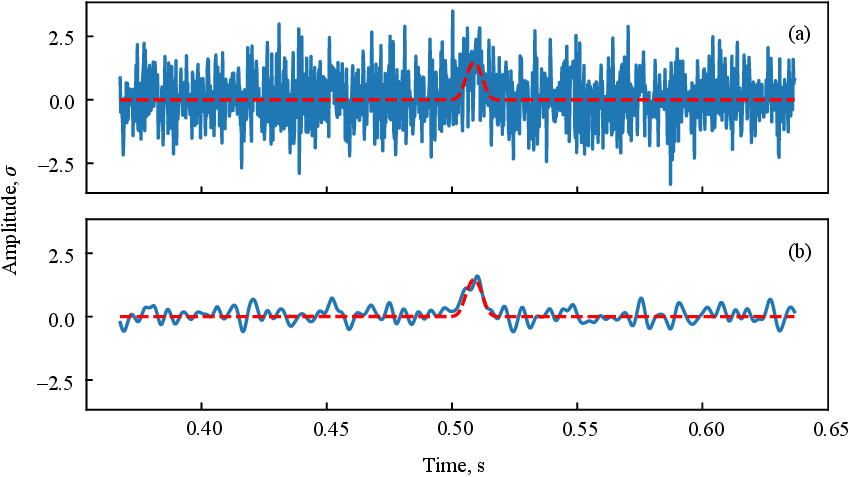}
\caption{Noise reduction using a low-pass Butterworth filter.
The red line represents the synthetic intrinsic time profile with known parameters. The top panel shows this profile injected with white noise ($\rm SNR = 1.5$), while the bottom panel displays the filtered output demonstrating the effective recovery of the pulse.}
\label{fig:filter}
\end{figure*}

The only parameter of the Butterworth filter that needs to be determined is its cutoff frequency, which separates the low and high temporal frequencies. This frequency can be found if the FWHM of the pulse time profile to be recovered is known. As it can be demonstrated by the Fourier transform, the frequency components corresponding to the signal effectively fall below $\rm 1/FWHM$~[Hz], where the $\rm FWHM$ of the pulse is given in seconds. This value can thus be adopted as the cutoff frequency of the filter.

When processing candidate events, we first measure the FWHM of the pulse from the original time series using the Bayesian inference approach described in the previous subsection. We then apply the Butterworth filter and remeasure the final pulse parameters from the data with the high-frequency noise suppressed.

It is worth noting that while any pulse could theoretically be submerged below the noise level by increasing the radiometer temporal resolution, the underlying statistical deviation remains preserved and can be found using sophisticated techniques such as neural networks. The subsequent noise reduction allows us to recover the highest possible SNR, independent of the chosen temporal resolution.

\section{DETECTED EVENTS}

Figures~3 and 4 present the candidate events observed on May 11, 2019, and June 25, 2019. The original observed time series are shown in the left panels of the figures, while the right panels present the Butterworth-filtered data. The applied noise reduction clarifies temporal profiles, revealing maxima the positions of which are consistent with the cold-plasma dispersion relation from Eq.~(\ref{eq:time_delay}). The observed amplitude variations are attributed to interstellar scintillation and nulling effects.

\begin{figure*}[ht!]
\includegraphics[width=0.97\linewidth]{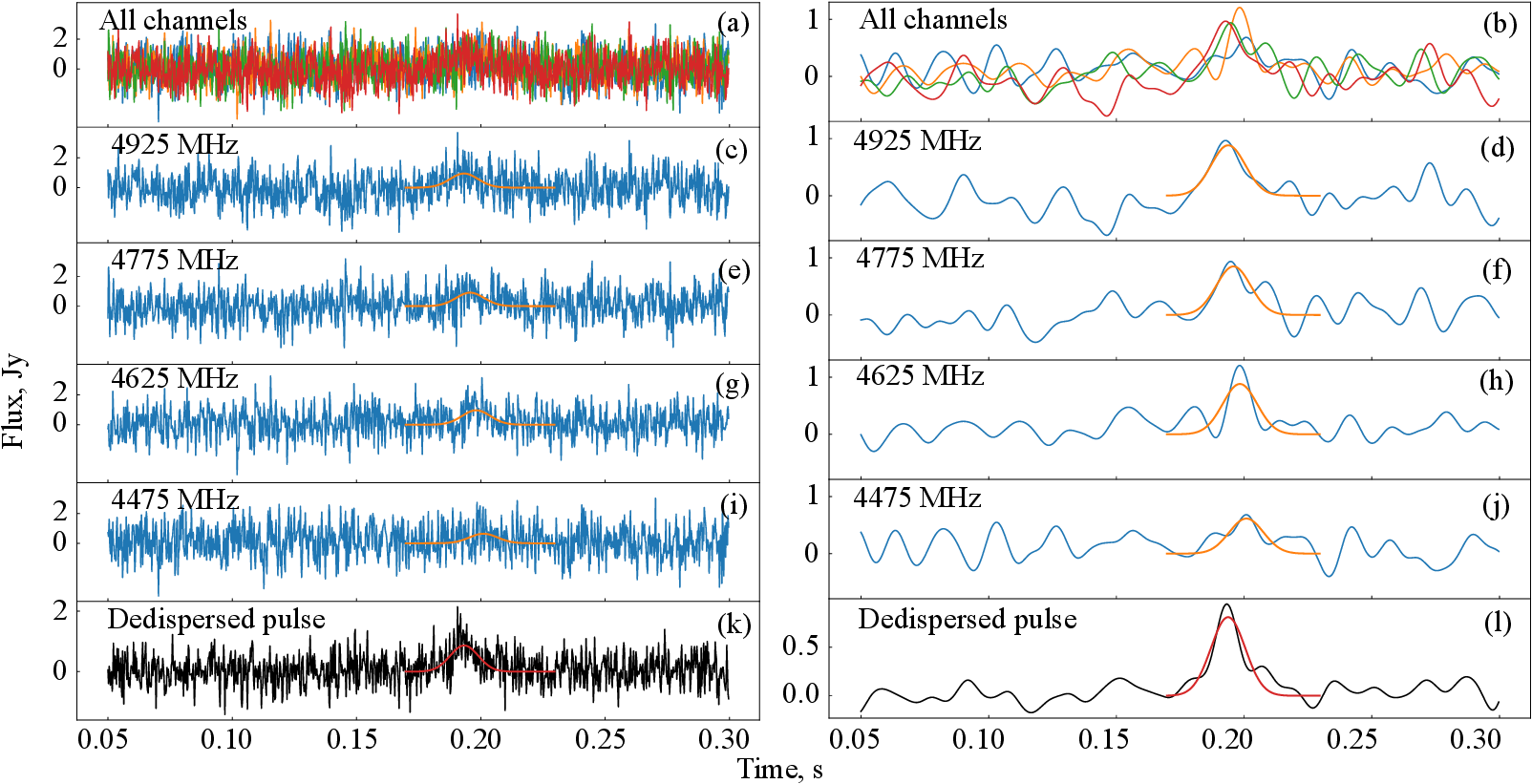}
\caption{Candidate event on 11~May~2019. The left panel presents the original observed record, the right panel shows the same record after applying the low-pass Butterworth filter.  In each panel the top subplot shows all the four-channel records combined, the lower subplots show the per-channel records, and the bottom subplot presents the dedispersed total pulse.}
\label{fig:bayes_fit_20190511}
\end{figure*}

\begin{figure*}[ht!]
\includegraphics[width=0.99\linewidth]{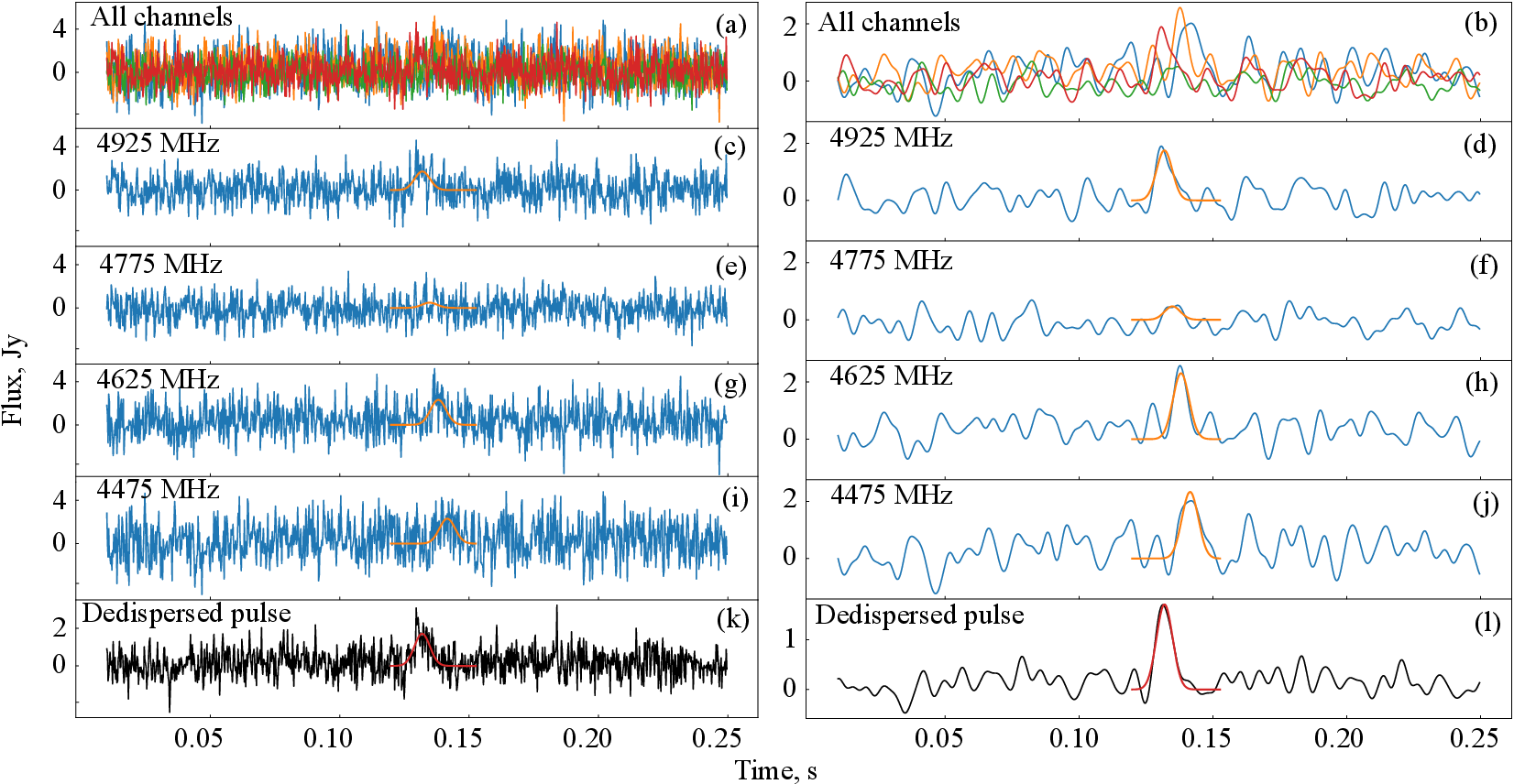}
\caption{Same as in Fig.~3, but for the candidate event on 25~June~2019.}
\label{fig:bayes_fit_20190625}
\end{figure*}

Figures~5 and~6 present the $\rm DM$--time diagrams, as well for both the original and noise-suppressed time series. Despite the inherently poor $\rm DM$ localization expected for a few-channel system, the candidate FRBs are clearly distinguishable from the background noise, exhibiting peak intensities at non-zero $\rm DM$ values---a defining characteristic of signals of cosmic origin. This effect is apparent in both the original and noise-suppressed time series. The red crosses in the figures indicate the best-fit parameters obtained obtained from the Bayesian fitting.

In summary, the observed deviations significantly exceed the noise floor, and their arrival times align closely with the cold-plasma dispersion relation. The recovered time profiles generally exhibit symmetric rise and decay phases consistent with a Gaussian-like morphology. The high-frequency noise suppression enhances the overall clarity of the profiles. The inter-channel amplitude variations is an expected effect caused by interstellar scintillation. Taken together, these characteristics allow us to confidently classify the events as strong FRB candidates.

\begin{figure*}[ht!]
\includegraphics[width=0.85\linewidth]{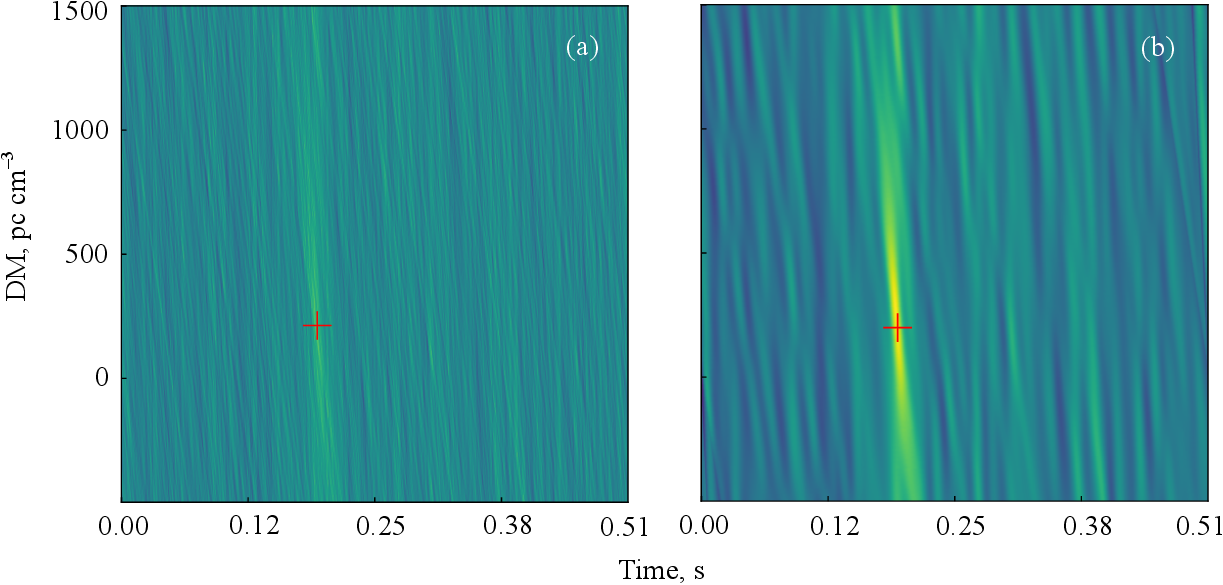}
\caption{Candidate event on 11 May 2019 represented in the DM--time domain. Left: the original record; right: the record after applying the low-pass Butterworth filter. The red cross designates the pulse position according to the best-fit model.}
\label{fig:dmt_20190511}
\end{figure*}

\begin{figure*}[ht!] \vspace{-2mm}
\includegraphics[width=0.85\linewidth]{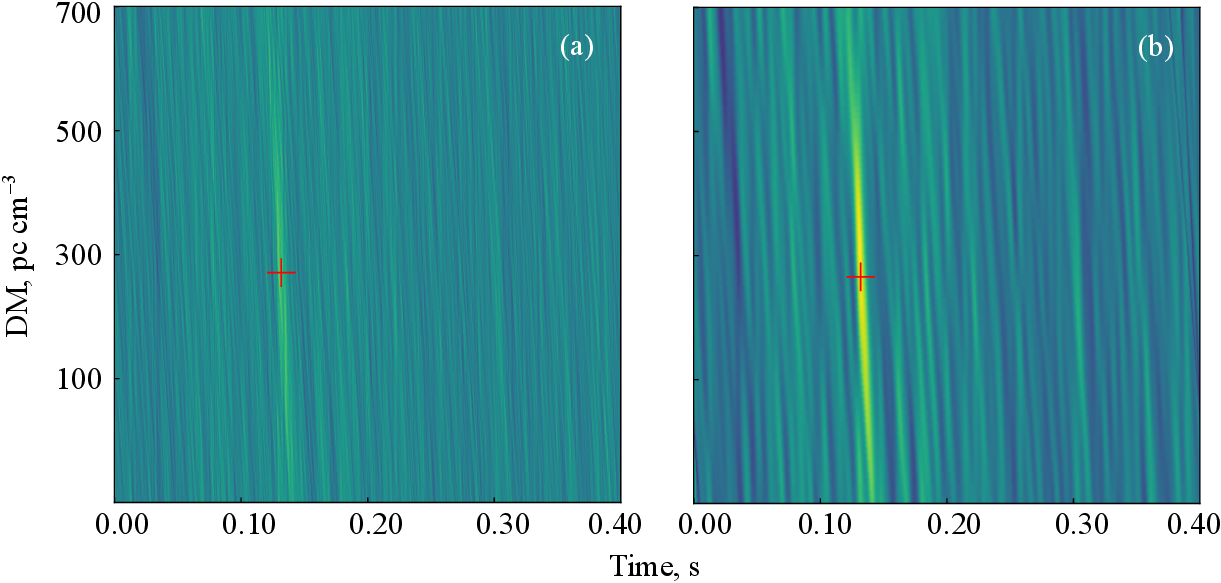}
\caption{Same as in Fig.~5, but for the candidate event on 25 June 2019.}
\label{fig:dmt_20190625}
\end{figure*}

\renewcommand{\baselinestretch}{0.75}
\setlength{\tabcolsep}{3pt}
\begin{table*}[]
\caption{Parameters of the events}\medskip
\label{tab:event_parameters}
\begin{tabular}{l|c|c}
\hline
\multicolumn{1}{c|}{\multirow{2}{*}{Parameter}}               & MJD58614.38257                     & MJD58659.62482  \\
			& 2019/05/11                        & 2019/05/25             \\
\hline
ICRS RA(2000)           & $ 07^{\rm h}\,31^{\rm m}\,42\fs73$ &  $17^{\rm h}\,37^{\rm m}\,58\fs31$ \\
ICRS Dec(2000)          & $+22^\circ\,03'\,54\farcs7$        &  $+05^\circ\,01'\,33\farcs42$       \\
$l$, deg                & 197.387                            & 29.258            \\
$b$, deg                & 18.985                             & 18.063            \\
Arrival time, UTC       & 09:10:54.138                       & 14:59:44.490      \\
DM, cm$^{-3}$\,pc       & $200\pm20$                         & $266\pm10$        \\
DM\,(MW), cm$^{-3}$\,pc    & 81                                 & 105               \\
DM\,(halo), cm$^{-3}$\,pc  & 30                                 & 30               \\
Redshift $z$              & $0.090\pm0.03$                     & $0.131\pm0.11$     \\
Width, ms               & $15.8\pm0.8$                       & $6.7\pm0.3$       \\
Fluence Jy\, ms         & $13.6\pm1.4$                       & $12.9\pm1.3$      \\
Peak flux density, Jy   & $0.81\pm0.04$                      & $1.7\pm0.1$       \\
SNR (after filter)      & 9.0                                & 8.3               \\
\hline
\end{tabular}
\end{table*}
\renewcommand{\baselinestretch}{1}

The parameters of the pulses, measured from Bayesian fitting as described in  Section~2.1, are given in Table~1. We have used the NE2001 model to estimate the Milky Way contribution, DM(MW), to the observed dispersion 
\citep{2002astro.ph..7156C}
and adopted \mbox{$\rm DM (halo) = 30$\,pc\,cm$^{-3}$} for the hot diffuse gas in the Galactic halo  (Liu et al., 2026).\citep{2026ApJ...998...15L}.
After that, the redshift is estimated from the well-known equation ${\rm DM\,[pc\,cm^{-3}]} = 1000\,z$ \citep{2019A&ARv..27....4P}. 

To determine the possible locations of the candidate events, we must estimate their error boxes in the $\rm{RA}_{\rm{J2000}}$ and $\rm{Dec}_{\rm{J2000}}$ coordinate system. Crucially, the observations were carried out using the antenna pointed toward the East (azimuth $A=90^\circ$). The telescope beam pattern is characterized by a wide vertical (elevation) size of $36^\prime$ and a narrow horizontal (azimuthal) size of $1^\prime$. To transform the beam pattern from the horizontal (local) coordinate frame to the equatorial coordinates (RA, Dec), one can utilize the following relations:
\begin{align}
\label{eq:RADEC}
\Delta \rm{RA}  &= \Delta h \, \sin{\rm PA} \pm 0.5  \phi \, \cos{\rm PA}, \\
\Delta \rm{Dec} &= \Delta h \, \cos{\rm PA} \pm 0.5  \phi \, \sin {\rm PA},
\end{align}
where $\rm{RA}(0)$ and $\rm{Dec}(0)$ are the nominal coordinates of the detected FRB events, $\rm PA$ is the position angle of the equatorial coordinate grid relative to the local horizon, and $\Delta h$ represents the angular offset from the beam center. The size of the beam pattern is defined by $\phi = 1^\prime$ and $\Delta h = \pm 18^\prime$.

The positions of the telescope beam pattern in both the horizontal and equatorial coordinate systems for the two events are illustrated in Fig.~7.

\section{DISCUSSION AND CONCLUSION}

We have presented the results of the high-frequency (4.7\,GHz) blind FRB searches conducted with the Western Sector of the \mbox{RATAN-600} radio telescope. The overall observing program since 2017 encompasses nine distinct annual surveys, including four complete surveys and five partial surveys (with coverage fractions ranging from 0.5 to 0.75). Operating as a transit instrument, the telescope provided continuous, round-the-clock monitoring of a $360^\circ \times 0\,.\!\!^\circ6$ sky strip. The antenna configuration, coupled with four independent radiometers, provided four separate beams on the sky at the same elevation but sequentially offset from one another by 10--15~s  in sidereal time or $2\,.\mkern-4mu^\prime1$ in azimuth. Primary data acquisition was performed with a temporal resolution of 245\,$\mu${s}.

\begin{figure*}[t]
\includegraphics[width=0.93\linewidth]{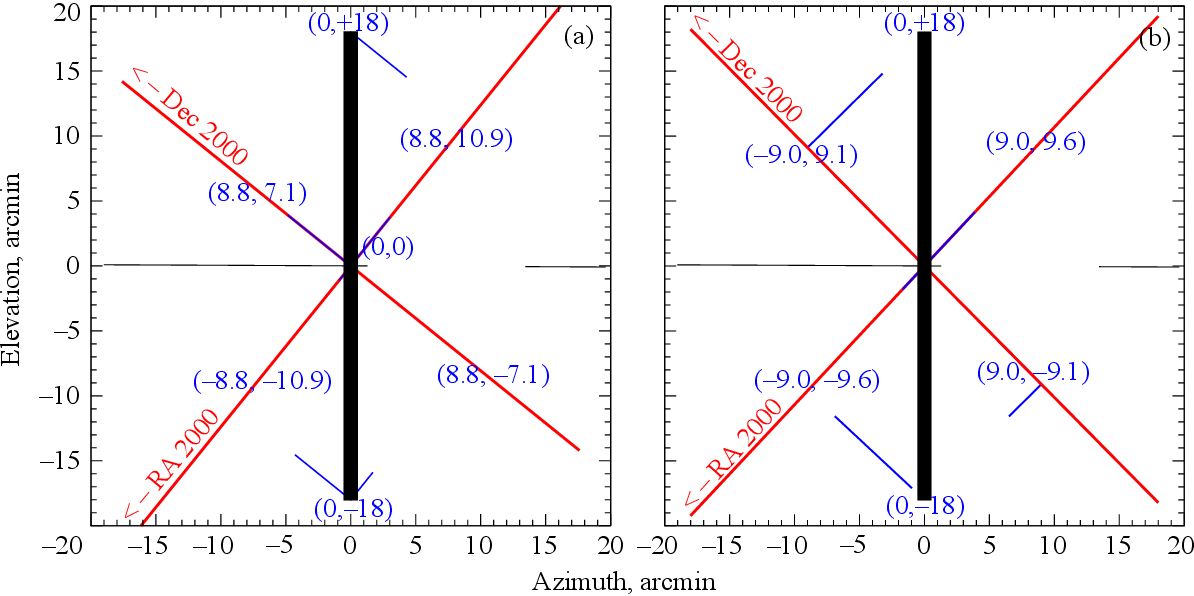}
\caption{Error boxes for the FRB candidate events detected on 11~May 2019 ($\rm{Dec}_{\rm{(J2000)}} = +22^\circ$,  panel (a)) and 25~June 2019 ($\rm{Dec}_{\rm({J2000)}} = +5^\circ$, panel (b)). The origin (null point) in each panel corresponds to the nominal coordinates of the respective event. The coordinate pairs indicate the spatial boundaries of the telescope beam pattern.}
\label{fig:errors}
\end{figure*}

The two prominent FRB candidates reported in this work were detected in the \mbox{2018--2019} observing epochs at declinations of $\rm{Dec}\approx +22^\circ$ and $+5^\circ$. The observed width of an FRB pulse ($w_{\rm obs}$) is generally produced by its intrinsic width ($w_{\rm int}$), cosmological redshift ($z$), interstellar scattering ($\tau_{\rm scat}$), and instrumental intra-channel dispersion smearing ($\Delta t_{\rm DM}$), such that \mbox{$w_{\rm obs} \approx w_{\rm int}(1+z) + \Delta t_{\rm DM}$}, given that scattering is negligibly small at 4.7\,GHz. For a channel bandwidth of 150\,MHz, the intra-channel dispersion smearing contributes approximately 2.8~ms (for $\rm DM = 266\,pc\,cm^{-3}$) and 2.1~ms (for \mbox{$\rm DM = 200\,pc\,cm^{-3}$}). These effects are incorporated in the calculation of the model profiles, and Table~1 gives the actual, intrinsic widths.

The flux densities of the pulses were calibrated using nearby bright point sources observed within the same drift scans. Crucially, because the exact trajectories of the sources across the telescope beams remain unconstrained, the measured peak fluxes and subsequent fluences (about 13~Jy~ms) must be treated as lower limits. Depending on the true offset from the beam centers, the actual flux densities could be higher by a factor of up to four.

Throughout the analysis of the records of this long-term program, the neural network processed approximately 750~billion 1-s data segments. After a rigorous filtering pipeline to eliminate false-positive RFI signals, followed by modeling the pulses using a Gaussian profile and low-pass Butterworth filtering, only these two highly significant events were recovered, each achieving a post-filtering signal-to-noise ratio ($\rm SNR$) exceeding 8. While the relatively large width of the second event (about 16~ms) is more characteristic of low-frequency FRBs, both bursts were detected at the very edge of our instrumental sensitivity threshold.

Ultimately, our results indicate that both the instantaneous field of view and the limited number of receiver channels of the current \mbox{RATAN-600} setup restrict the detection rate of cosmic bursts. This is consistent with accumulating observational evidence suggesting that FRB emission often features a relatively narrow bandwidth clustered around \mbox{800--1400\,MHz}, with spectral energy distributions decaying noticeably at frequencies above 3\,GHz. Although repeating FRBs might possess broader emission spectra, their volumetric population is substantially smaller than that of non-repeating sources.

Obtaining the luminosity distances $D_l$ from the redshifts and the commonly accepted $\Lambda$CDM model parameters, we can estimate the isotropic radio luminosities using the formula:
\begin{equation}
 L_{\rm rad}  =  \cfrac{4 \pi D_l^2 S_\nu \nu }{1+z}.
\end{equation}
The two estimates are equal to $7.2 \times 10^{41}$~erg\,s$^{-1}$  and $3\times 10^{42}$~erg\,s$^{-1}$, respectively. Thus, these calculations yield event energies of $10^{40}$~erg and $2\times 10^{40}$~erg, respectively. These values are well within the total broad  range of estimates for known FRBs, while they are somewhat higher than the average value of $10^{39}$~erg. It's not surprising that both FRB energy estimates are similar and somewhat brighter than average:
only sufficiently bright events could be detected given the flux limit of our survey.

We can estimate the rate of high-frequency FRBs based on the two our detections by scaling the surveyed sky area, which yields  $41\,253/1500 = 27.5$ events per year per whole sky. In the Commensal Real-time ASKAP Fast Transients survey (CRAFT) at 1.4\,GHz, a rate of 12.7 per day for FRBs brigher than 57\,Jy\,ms is given \citep{2019PASA...36....9J}. 
Such a threshold corresponds to 1\,Jy at 4.7\,GHz if we take a mean FRB width of 10\,ms. This realistic rate is equal to 4635~events per year and per sky. We see that our rate estimate is  nearly two orders of magnitude lower that derived from the CRAFT data. The most likely reason of such a difference is a physically narrow emission frequency range for non-repeating bursts, a point that has been widely discussed in the literature. That is, it cannot be ruled out that the primary frequency range of the brightest FRBs lies within $1000\pm500$\,MHz, while at frequencies above and below this window, FRB brightness drops sharply, leaving only rare events detectable.

The blind survey is ongoing in 2026, and we hope achieve new FRB detections. To optimize future transient searches at \mbox{RATAN-600}, we plan to transit to a lower central frequency (2\,GHz), utilizing a multi-channel spectrometer. Concurrently, the existing 4.7\,GHz receiving complex will be repurposed for targeted, rapid-alert monitoring of known repeating FRBs with an upgraded sampling rate of 50~$\mu$s.

\section*{ACKNOWLEDGMENTS}
We are grateful to the technical staff of \mbox{RATAN-600} for the maintenance of the Western Sector antenna. Observations with the SAO~RAS telescopes are supported by the Ministry of Science and Higher Education of the Russian Federation. The renovation of telescope equipment is currently provided within the national project ''Science and Universities.\!''

\section*{FUNDING}
This work was carried out as part of the SAO~RAS state assignment, approved by the Ministry of Science and Higher Education of the Russian Federation.


\section*{AUTHOR CONTRIBUTIONS}
The co-authors contributed to the work in the following capacities:
\begin{list}{}{
\setlength\leftmargin{4mm} \setlength\topsep{0mm}
\setlength\parsep{0mm} \setlength\itemsep{0.mm} }
    \item[$\bullet$] {S.T.} is the principal investigator who initiated and led the ``FRB with RATAN-600'' project;
     \item[$\bullet$] {P.T.} served as the lead radiometer scientist who proposed, designed, and commissioned the four-channel radiometer complex;
     \item[$\bullet$] {D.K.} was the primary developer of the FRB search, detection, and modeling pipeline utilizing neural networks;
     \item[$\bullet$] {N.B.} acted as the principal observer and managed the coordination of the survey observations.
\end{list}

\end{document}